\documentclass[twocolumn,aps,prl,floatfix,showpacs]{revtex4}
\usepackage{bm}
\usepackage{amssymb,amsmath}
\usepackage{dcolumn}
\usepackage{graphicx}
\usepackage{bbold}
\usepackage{booktabs}
\usepackage{color}
\usepackage[colorlinks,bookmarks=false,citecolor=blue,linkcolor=blue,urlcolor=blue]{hyperref}

\def\ahat{\hat{a}}
\def\bhat{\hat{b}}

\def\Phihat{\hat{\Phi}}
\def\rhohat{\hat{\rho}}

\def\Mhat{\hat{M}}
\def\Nhat{\hat{N}}
\def\calU{{\cal U}}
\def\calD{{\cal D}}

\def\calG{{\cal G}}
\def\calO{{\cal O}}

\begin{document}
\title{Emergent Radiation in an Atom-Field System at Twice-Resonance}
\author{Brijesh Kumar}
\email{bkumar@mail.jnu.ac.in}
\affiliation{School of Physical Sciences, Jawaharlal Nehru University, New Delhi 110067, India.}
\date{October 25, 2008} 
\begin{abstract}
A two-level atom interacting with a single mode of quantized electromagnetic radiation is discussed using a representation in which the atom and the radiation are unified into a {\em new} canonical radiation. At the {\em twice-resonance}, when the frequency of the original radiation is twice the atomic transition frequency ($\omega=2\epsilon$), the {\em emergent} unified field in the non-interacting atom-field system resembles a free radiation of frequency $\epsilon$. This free emergent radiation is further shown to exist in the presence of an interaction which looks similar to the atom-field interaction in the dipole approximation. The one-photon correlation and the population inversion are discussed as the possible means of observing the emergent radiation. The entanglement properties of the emergent radiation are also discussed.
\end{abstract}
\maketitle
The matter-radiation interaction problems are of central importance to all branches of physics. A typical problem of this kind, in quantum optics for example, concerns the interaction of light with an atom in the long wavelength (dipole) approximation~\cite{Scully_Zubairy,Science_CavityQED}. In a useful simplification of the atom-field problem, the atom is often approximated as an effective two-level system. In a fully quantum-theoretic  formulation, the light is treated as a quantized radiation-field. The corresponding atom-field Hamiltonians, for example the Dicke maser model~\cite{Dicke} or the Rabi model~\cite{Rabi}, form the basis of understanding for a number of physical phenomena, such as the cooperative superradiant emission~\cite{Dicke}, or the oscillations of the population inversion and their collapse and revival~\cite{collapse_cummings,collapse_eberly,collapse_exp}.

 In this Letter, we consider a two-level atom interacting with a single mode of quantized radiation, and predict the emergence of a {\em free} unified radiation at the twice-resonance. The term twice-resonance refers to the condition when the frequency of the original radiation is twice the atomic transition frequency. The present discussion is based on, and inspired by, a representation in which a two-level atom (Pauli operators) and a single mode of quantum radiation (Bose operators) are unified into a {\em single} Bose field~\cite{bk_rep}. The existence of this free emergent radiation, whose frequency is same as the atomic transition frequency, is first demonstrated in the non-interacting case. It is then shown to exist in an interacting system, which is similar to the atom-photon model in the dipole approximation. We discuss the time dependence of the one-photon correlation and the population inversion in simple situations, to illustrate the possibility of observing this effect. We also present a discussion on the atom-field entanglement in the emergent radiation.

Let $|g\rangle$ and $|e\rangle$ denote the ground and excited states of a two-level atom, respectively. Let $\epsilon$ be the energy-difference of the two atomic levels and $\bar{\epsilon}$ be their average. The Hamiltonian of such an isolated atom can be written as: $H_{atom}=\bar{\epsilon}~\mathbb{1}+\frac{\epsilon}{2}~\sigma^z$, where $\mathbb{1}$ is the identity operator and $\sigma^z=|e\rangle\langle e| - |g\rangle\langle g|$ is the $z$-component of the Pauli operator, defined in the atomic Hilbert space. Physically, $\sigma^z$ measures the population inversion, that is, the difference of the probabilities of finding the atom in the excited and the ground states. The transition between the atomic levels is described in terms of the other two Pauli operators, $\sigma^+=|e\rangle\langle g|$ and $\sigma^-=|g\rangle\langle e|$, which excite or de-excite the atom, respectively. Together, $\sigma^z$ and $\sigma^\pm$ satisfy the usual spin-$1/2$ operator algebra. 

A single mode of the quantized electromagnetic radiation is described in terms of the Bose operators, $\bhat$ and $\bhat^\dag$, which annihilate and create a photon, respectively. These operators act in the Fock space of the photons, spanned by the basis, $\{ |m\rangle,~m=0,1,\dots,\infty \}$, such that $\bhat~|m\rangle=\sqrt{m}~|m-1\rangle$ and $\bhat^\dag~|m\rangle=\sqrt{m+1}~|m+1\rangle$. The ket $|m\rangle$ denotes the $m$-photon state for the given mode of the radiation. The Hamiltonian of such a single-mode quantized radiation-field can be written as: $H_{field}=\omega~(\bhat^\dag\bhat+\frac{1}{2})$, where $\bhat^\dag\bhat$ is the photon number operator, that is, $\bhat^\dag\bhat~|m\rangle=m~|m\rangle$, and $\omega$ is the photon energy ($\hbar=1$). 

The interaction of such a two-level atom with a single mode of quantized radiation can be described, in the dipole approximation, by $V_{dipole}=(\bhat^\dag + \bhat)\sigma^x$, where $\sigma^x=\sigma^+ +\sigma^-$ measures of the dipole moment of the atom, and the electric field of the radiation is proportional to $\bhat^\dag+\bhat$. The simplest description of the matter-radiation problem can therefore be carried out in terms of the Hamiltonian, $H_{dipole}=H_{field}+H_{atom} +gV_{dipole}$, where $g=d\sqrt{\omega/2\varepsilon_0V}$ is the dipole-radiation coupling ($d=$ atomic dipole matrix element)~\cite{Scully_Zubairy}. This model has been of fundamental interest to the studies in quantum optics and magnetic resonance~\cite{Rabi}. In the rotating wave approximation due to Jaynes and Cummings, for $\omega$ close to $\epsilon$, the $ V_{dipole}$ is approximated by $V_{JC}=(\bhat^\dag\sigma^- + \bhat\sigma^+)$, where the faster processes, $\bhat^\dag\sigma^+ + \bhat\sigma^-$, in $V_{dipole}$ have been dropped. This simplification results in the exactly solvable Jaynes-Cummings (JC) model~\cite{JC1}, which is an extensively studied model in quantum optics. The atom-field interaction in the present study is of a different type, but has a close similarity to the dipole interaction.

Recently, we have introduced a new representation in which the radiation operators, $\bhat$ and $\bhat^\dag$, and the atomic operators, $\sigma^z$ and $\sigma^\pm$, are unified into a single canonical Bose operator~\cite{bk_rep}. Let $\ahat^\dag$ be the creation operator of the unified boson. Then, according to our representation, 
\begin{equation}
\ahat^\dag =\sqrt{2}\left[\sqrt{\bhat^\dag \bhat+\frac{1}{2}}~\sigma^+ + \bhat^\dag \sigma^- \right]
\label{eq:bose_rep}
\end{equation}
It can be checked that Eq.~(\ref{eq:bose_rep}) satisfies the necessary bosonic commutation relations. The operator $\ahat^\dag$ acts on the new Fock states, $\{| n \rangle,~n=0,1,\dots,\infty\}$, which are defined as: $|n=2m\rangle = |m\rangle\otimes|g\rangle$ and $|n=2m+1\rangle = |m\rangle\otimes|e\rangle$. The representation in Eq.~(\ref{eq:bose_rep}) is based on this definition of the $|n\rangle$ kets (please refer to the Appendix B of Ref.~\cite{bk_rep} for details).  The corresponding inverse representation is  given by the following equations.
\begin{eqnarray}
\sigma_z &=& -\cos{(\pi\ahat^\dag\ahat)} :=-\hat{\chi} \label{eq:sigmaz_bose}\\
\sigma^+ &=& \frac{1-\hat{\chi}}{2}~\frac{1}{\sqrt{\Nhat}}~\ahat^\dag \label{eq:sigmap_bose}
\end{eqnarray}
and
\begin{equation}
\bhat^\dag = \frac{\ahat^\dag\ahat^\dag}{\sqrt{2}}\left(\frac{1-\hat{\chi}}{2}\frac{1}{\sqrt{\Nhat+2}} +  \frac{1+\hat{\chi}}{2}\frac{1}{\sqrt{\Nhat+1}} \right)
\label{eq:b_bose}
\end{equation}
where $\hat{\chi}$ can also be written as $(-)^{\Nhat}$, and $\Nhat=\ahat^\dag\ahat$ is the number operator of the unified boson. This representation of $\bhat^\dag$ is consistent with the fact that changing the number of $b$-bosons by 1 changes the number of $a$-bosons by 2. Besides, it satisfies the bosonic commutations, and commutes with $\sigma_z$ and $\sigma^\pm$, as defined in Eqs.~(\ref{eq:sigmaz_bose}) and~(\ref{eq:sigmap_bose}). 

We thus have a canonical description of  the original radiation and the atom in terms of the  unified boson. We will refer to this unified boson as the emergent or unified radiation. Through the inverse transformation, we can now convert any problem of a two-level atom interacting with a single-mode quantized radiation to an equivalent problem of the unified radiation. In general, this will be a highly non-linear problem in terms of the unified boson. Therefore, the corresponding energy eigenstates are not expected to be like that of a free electromagnetic radiation (the hallmark of a free radiation is the presence of equidistant successive photon states like a simple harmonic oscillator). In other words, the emergent radiation in an arbitrary atom-field system will be {\em distinguishable} from a normal free radiation. However, below we present a simple situation, in which the emergent radiation completely resembles a free radiation.
\begin{figure}[t]
   \centering
   \includegraphics[width=8.5cm]{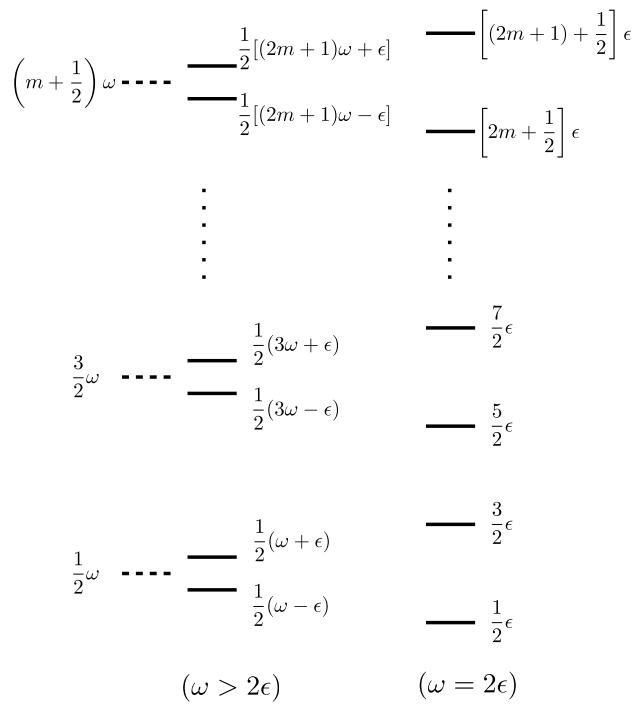}
   \caption{Energy-level diagram (in the non-interacting case) of a quantized single-mode radiation and a two-level atom. For $\omega=2\epsilon$, the states of atom-field system are indistinguishable from that of a radiation-only system. Hence, the twice-resonant case ($\omega=2\epsilon$) presents an emergent radiation.}
   \label{fig:energy_levels}
\end{figure}

We start in the reverse order by considering the free-field Hamiltonian for the unified radiation. Upto an overall factor of the energy, it has the following form.
\begin{equation}
\ahat^\dag\ahat+\frac{1}{2}= 2\left(\bhat^\dag\bhat +\frac{1}{2}\right) + \frac{1}{2}\sigma^z \label{eq:atom_field}
\end{equation}
The right-hand side of the above equation shows the corresponding Hamiltonian in terms of the original atom-field variables. It just happens to be a non-interacting atom-radiation problem at precisely the twice-resonance. It also implies that the ``photon'' energy of the emergent radiation is same as the atomic transition frequency. Thus, Eq.~(\ref{eq:atom_field}) presents a special atom-field system, in which the unified radiation emerges as ``free''. With the benefit of hindsight, now we can directly demonstrate the existence of this free emergent radiation without evoking the representation. In Fig.~\ref{fig:energy_levels}, we  present how, at the twice-resonance, the non-interacting atom-photon system is {\em indistinguishable} from a free radiation with photon energy $\epsilon$. Demanding that the successive eigenstates be equally spaced leads to the condition, $\omega=2\epsilon$~\cite{fnote_lomega}.

In experimental terms, it suggests that, if there is a two-level atom inside a cavity with precisely twice-resonant radiation, then in principle, one can not distinguish this system from another cavity with only the radiation of frequency $\epsilon$. Or, if there is a cavity with forbidden $\epsilon$-frequency mode (but allowed $2\epsilon$ modes), then it will actually sustain or exhibit an $\epsilon$-frequency radiation in the presence of a right atom. A test atom, with transition frequency close to $\epsilon$, will undergo Rabi oscillations inside such a cavity. The emergence of a free unified radiation in the non-interacting system, however, presents a trivial case. In reality, no atom is free from the interaction with the radiation. Therefore, it is important to ask whether this emergent radiation will survive in an interacting atom-photon system, or not~\cite{fnote_g}.

To further investigate the emergent radiation in the presence of an atom-photon interaction, we introduce such modifications in the free-field Hamiltonian of the unified radiation [Eq.~(\ref{eq:atom_field})] that the equidistant character of its eigenstates  survives. Clearly, the following general Hamiltonian fulfills our objective.
\begin{equation}
H = \epsilon\left(\ahat^\dag\ahat + \frac{1}{2}\right) + \xi\left(\ahat^\dag+\ahat\right)+\eta\left(\ahat^\dag\ahat^\dag + \ahat\ahat\right)
\label{eq:H_aboson}
\end{equation}
It is an exactly solvable model with a displacement term ($\propto \xi$) and the pairing ($\propto \eta$). This $H$ is diagonalized by the unitary transformations, $\calD(x)=e^{-x(\ahat^\dag-\ahat)}$ and $\calU(\theta)=e^{-\theta(\ahat^\dag \ahat^\dag - \ahat \ahat)}$, where $x=\frac{\xi}{\epsilon+2\eta}$ and $\theta=\frac{1}{4}\tanh^{-1}(\frac{2\eta}{\epsilon})$. For every ket $|n\rangle$ of the $a$-boson, the exact eigenstate of $H$ is given by $|\psi_n\rangle=\calD(x)\calU(\theta)|n\rangle$, with an eigenvalue, $E_n=  \sqrt{\epsilon^2-4\eta^2}~(n+\frac{1}{2})-\frac{\xi^2}{(\epsilon+2\eta)}$.
While $\xi$ uniformly lowers the eigenvalue for each $n$,  the $\eta$ also renormalizes the frequency of the unified radiation. However, the basic structure of the eigen-spectrum of $H$ remains like that of a free radiation [that is, $E_n\propto (n+\frac{1}{2})$]. The emergent radiation eigenstate, $|\psi_n\rangle$, is now an {\em entangled} state of the atom and the original radiation, unlike in the non-interacting case, where it is not entangled.

By using Eq.~(\ref{eq:bose_rep}), this $H$ can be converted to the following atom-field problem.
\begin{eqnarray}
H&=&2\epsilon H_b + \frac{\epsilon}{2}\sigma^z + \xi\sqrt{2}\left(\bhat^\dag\sigma^-+\bhat\sigma^+ +\sqrt{H_b}~\sigma^x\right) \nonumber \\
&&+\eta\left\{\sqrt{H_b}\left[\left(\bhat^\dag+\bhat\right)+\sigma^z\left(\bhat^\dag-\bhat\right)\right]+h.c.\right\}, \label{eq:H_bsigma}
\end{eqnarray}
where $H_b=\bhat^\dag\bhat+\frac{1}{2}$. According to this form, the atom and the radiation interact via different terms arising due to $\xi$ and $\eta$. The atom-field interactions in Eq.~(\ref{eq:H_bsigma}) look rather complicated.
However, note the interaction, $\bhat^\dag\sigma^- + \bhat\sigma^+$, arising due to $\xi$. It is the well-known Jaynes-Cummings interaction, $V_{JC}$. To keep things simple, we therefore set $\eta=0$ for the rest of our discussion. A non-zero $\xi$ is sufficient to generate the desired atom-photon interaction.

We now have a sufficiently non-trivial but simple model in $H$ for $\eta=0$, which preserves the free-field character of the emergent radiation at the twice-resonance. The corresponding atom-photon Hamiltonian is written as: 
\begin{equation}
H_0=2\epsilon H_b + \frac{\epsilon}{2}\sigma^z + \xi\sqrt{2}(\bhat^\dag\sigma^- + \bhat\sigma^+ +\sqrt{H_b}~\sigma^x).
\label{eq:H_0}
\end{equation}
Besides the physical $V_{JC}$, the $H_0$ also has another piece to the interaction, that is $\sqrt{H_b}~\sigma^x$, which has not been encountered before. To check whether this new interaction can be ignored within the rotating wave approximation, we compare its time dependence (in the interaction picture) with that of the JC interaction. While the time dependence of $\sqrt{H_b}~\sigma^x$ is $\sqrt{H_b}~(\sigma^+ e^{i\epsilon t} + \sigma^- e^{-i\epsilon t})$, the $V_{JC}$ varies as $\bhat^\dag\sigma^-e^{i\epsilon t} + \bhat\sigma^+ e^{-i\epsilon t}$. Since both have similar time dependences, we can not drop either of the two on the grounds of being faster. Moreover, we need both of these for maintaining the free-field character of the emergent radiation. Therefore, it would be nice to have some physical understanding for $\sqrt{H_b}~\sigma^x$, so that it can at least formally be considered realistic. Surprisingly, we can show that $\sqrt{H_b}~\sigma^x$ is a unitary transformed version of the interaction, $\bhat^\dag\sigma^+ +\bhat\sigma^-$, which gets dropped from the $V_{dipole}$ in the rotating wave approximation.

To establish this unitary connection, consider the original radiation in the $number$-$phase$ representation. That is, $\bhat^\dag=\sqrt{\Mhat}~e^{-i\Phihat}$ and $\bhat=e^{i\Phihat}\sqrt{\Mhat}$, where the Hermitian operators $\Mhat$ and $\Phihat$ denote the number and the phase operators, respectively. Clearly, $\Mhat=\bhat^\dag \bhat $ and $e^{i\Phihat} \Mhat e^{-i\Phihat} = \Mhat + 1$. The latter also implies $[\Mhat,\Phihat]=i$. Hence, $\Mhat$ and $\Phihat$ are a pair of conjugate operators, like the position and momentum. In this representation, $\bhat^\dag\sigma^+ +\bhat\sigma^-$ is equal to $\sqrt{\Mhat}~e^{-i\Phihat}\sigma^+ + \sigma^- e^{i\Phihat}\sqrt{\Mhat}$. This expression prompts us to absorb the ``phase operator'' into $\sigma^\pm$. To achieve this, we introduce a unitary transformation, $\calU_{\Phihat}=e^{-\frac{i}{2}\sigma^z\Phihat}$, such that  $\calU^\dag_{\Phihat}~\sigma^\pm~\calU_{\Phihat} = \sigma^\pm~e^{\pm i\Phihat}$. Interestingly, under this transformation, we get 
\begin{equation}
\calU^{\dag}_{\Phihat}(\bhat^\dag\sigma^+ +\bhat\sigma^-)\calU_{\Phihat} = \sqrt{\Mhat +\frac{1}{2}}\;\sigma^x = \sqrt{H_b}~\sigma^x
\end{equation}
This is a beautiful result. It establishes a meaningful relation between the interaction in $H_0$, which is $\sqrt{H_b}~\sigma^x+ (\bhat^\dag\sigma^- + \bhat \sigma^+)=\calU^{\dag}_{\Phihat}(\bhat^\dag\sigma^+ +\bhat\sigma^-)\calU_{\Phihat} + (\bhat^\dag\sigma^- + \bhat \sigma^+)$, and the $V_{dipole}$, that is, $(\bhat^\dag\sigma^+ +\bhat\sigma^-) + (\bhat^\dag\sigma^- + \bhat \sigma^+)$.

Now we discuss some physical results within $H_0$, concerning the observation of the emergent radiation. We first compute the time-dependent correlation function of the electric-field, $E_b(t)$, of the original radiation. Since $E_b(t)\propto \bhat^\dag(t) + \bhat(t)$, the corresponding one-photon correlation $\langle E_b(t)E_b(0)\rangle$ is proportional to $\calG(t)$, where
\begin{equation}
\calG(t)=\left\langle\left[\bhat^\dag(t) + \bhat(t)\right]\left[\bhat^\dag(0) + \bhat(0)\right]\right\rangle.
\label{eq:g_t}
\end{equation}
Here, $\bhat(t) = e^{i H_0t}\bhat e^{-i H_0t}$, and for any operator $\hat{X}$, the expectation $\langle \hat{X} \rangle$ is calculated as $tr\{\rhohat \hat{X} \}$, where $\rhohat$ is a given density matrix. Since $\calD^\dag(x) H_0\calD(x)=\tilde{H}_0=\epsilon (\ahat^\dag\ahat +\frac{1}{2}) -\frac{\xi^2}{\epsilon}$ for $x=\xi/\epsilon$, we also transform $\bhat^\dag$ to $\bhat^\dag(x)=\calD^\dag(x) \bhat^\dag \calD(x)$, and similarly $\rhohat$ to $\rhohat(x)$. Therefore, $\calG(t)=tr\{\rhohat(x) [\bhat^\dag(x,t) + \bhat(x,t)][\bhat^\dag(x,0) + \bhat(x,0)]\}$, where $\bhat(x,t)=e^{i\tilde{H}_0t}\bhat(x) e^{-i\tilde{H}_0t}$. In order to demonstrate how, in principle, the emergent radiation will manifest itself through $\calG(t)$, we discuss two limiting cases: the weak- $(x \ll 1)$ and the strong-coupling $(x \gg 1)$. For simplicity, we may  take $\rhohat$ to be either the equilibrium density matrix, $e^{-\beta H_0}/tr\{e^{-\beta H_0}\}$, or the pure eigenstates of $H_0$, that is  $|\psi_n\rangle\langle\psi_n|$. For these two choices, $\rhohat(x)=e^{-\beta\tilde{H}_0}/tr\{e^{-\beta\tilde{H}_0}\}$ and $|n\rangle\langle n|$, respectively.

In the weak-coupling limit, we write $\bhat^\dag(x)+\bhat(x)$ as:
\begin{eqnarray}
\bhat^\dag(x) + \bhat(x) &\approx& (\bhat^\dag + \bhat) -x\sqrt{2}\left\{\left(\sigma^+ +\sigma^-\right) \right. \nonumber\\
&& \left. - \left[\sqrt{H_b},\bhat^\dag+\bhat\right]\left(\sigma^+ -\sigma^-\right)\right\} \label{eq:bx_weak},
\end{eqnarray}
where the terms of $\calO(x^2)$ have been ignored. In this limit, we get the following expression for $\calG(t)$.
\begin{eqnarray}
\calG(t)_{\xi\ll\epsilon} &\approx& \langle \bhat^\dag\bhat\rangle e^{i2\epsilon t} + \langle \bhat\bhat^\dag\rangle e^{-i2\epsilon t} + 2x^2 \left\{e^{i\epsilon t} \left[\langle\sigma^+\sigma^-\rangle\right. \right. \nonumber \\
&& \left.+B_1\langle\sigma^-\sigma^+\rangle\right] + e^{-i\epsilon t} \left[\langle\sigma^-\sigma^+\rangle + B_2\langle\sigma^+\sigma^-\rangle\right]\nonumber \\
&& \left. +B_1\langle\sigma^+\sigma^-\rangle e^{i3\epsilon t} +B_2\langle\sigma^-\sigma^+\rangle e^{-i3\epsilon t}\right\}, \label{eq:calG_weak}
\end{eqnarray}
where $B_1=\langle[\sqrt{H_b},\bhat^\dag][\bhat,\sqrt{H_b}]\rangle$, and $B_2=\langle[\sqrt{H_b},\bhat][\bhat^\dag,\sqrt{H_b}]\rangle$. The presence of the $e^{\pm i\epsilon t}$ terms in Eq.~(\ref{eq:calG_weak}) gives a clear indication of the dynamically generated free emergent radiation in the system. It is fascinating that, in a system with the radiation of frequency $2\epsilon$, the {\em fundamental note} of frequency $\epsilon$ appears due to the interaction. It is as if the original photon has split into two new photons with half the energy. Even for an arbitrary value of $x$, the $\calG(t)$ will only have the terms of frequency $\epsilon$ and its higher harmonics. 

In the strong-coupling limit, we can write
\begin{equation}
\bhat^\dag(x)+\bhat(x) \approx \frac{1}{\sqrt{2}}\left[ 2x - (\ahat^\dag+\ahat) + \frac{3}{4x}(\ahat^\dag - \ahat)^2 \right] \label{eq:bx_strong}
\end{equation}
where the terms of $\calO(1/x^2)$ have been ignored. The terms containing $\hat{\chi}(x)$, which roughly fall as $e^{-2x^2}/x$ for large $x$, have been completely ignored. In this case, the correlation function can be written as:
\begin{eqnarray}
\calG(t)_{\xi\gg\epsilon} &\approx& c_0 + \frac{1}{2}\left[\langle\ahat\ahat^\dag\rangle e^{-i\epsilon t} + \langle\ahat^\dag\ahat\rangle e^{i\epsilon t}\right] + \nonumber \\
&& \frac{9}{32x^2}\left[\langle \ahat\ahat\ahat^\dag\ahat^\dag\rangle e^{-i2\epsilon t} + \langle \ahat^\dag\ahat^\dag\ahat\ahat\rangle e^{i2\epsilon t}\right] \label{eq:calG_strong}
\end{eqnarray}
where $c_0=2x^2 -3[\langle\ahat^\dag\ahat\rangle +\frac{1}{2}] + \dots$,  is a constant term. The terms, $e^{\pm i\epsilon t}$, corresponding to the free emergent radiation again appear in the correlation function. In fact, a striking feature of Eq.~(\ref{eq:calG_strong}) is that, the $e^{\pm i2\epsilon t}$ terms of the original radiation, are sub-leading (in powers of $1/x$) compared to the emergent radiation terms. Therefore, the strong-coupling case of $H_0$ presents an {\em inverse} of the second-harmonic generation effect. This new effect may be called as the {\em half-} or sub-harmonic generation. 

Now, we briefly discuss the time evolution of the population inversion. For simplicity, we consider only the following choices for the initial state: $ |n=0\rangle= |m=0\rangle\otimes |g\rangle$,  and $ |n=1\rangle=|m=0\rangle\otimes |e\rangle$. In both cases, the original radiation is in the vacuum state. The exact population inversion in the two cases is: $W_0(t)=\langle0|\sigma^z(t)|0\rangle=-e^{-4x^2(1-\cos{\epsilon t})}$, and similarly, $W_1(t)=[1-8x^2(1-\cos{\epsilon t})] e^{-4x^2(1-\cos{\epsilon t})}$. Both $W_0$ and $W_1$ oscillate in time with a frequency $\epsilon$. The oscillation frequency, however, is independent of the atom-photon coupling, unlike in the Rabi oscillations. For $x\ll1$, both $W_0$ and $W_1$ make simple sinusoidal oscillations. The behavior for $x\gg1$ changes to periodic pulsing: $W_0=-\delta_{t=t_l}=-W_1$, where $t_l=\frac{2\pi}{\epsilon}l$, and $l$ is an integer. That is, both $W_0$ and $W_1$ remain zero except at $t_l$ when the atom only momentarily recovers (and loses) the initial state. The zero population inversion implies an atomic coherent state, with an equal probability for the atom to be in $|g\rangle$ and $|e\rangle$. This result offers a method for preparing the atomic coherent states by means of realizing the emergent radiation. It may be of interest to the coherent atom optics. The above features in a population inversion measurement could be taken as the signatures of the emergent radiation.

Finally, we discuss entanglement in the emergent radiation state. We first introduce a simple measure of entanglement for a bipartite system in a pure state. Let $X$ and $Y$ denote the two subsystems, and $\rhohat$ be the density operator of the full system. In a pure state, $\rhohat$ is just a projection operator. For this case, we define a measure of entanglement as: $\mathcal{E}=1-tr^{ }_X \rhohat_X^2 =1-tr^{ }_Y \rhohat_Y^2 $, where $\rhohat^{ }_X=tr^{ }_Y \rhohat $ and $\rhohat^{ }_Y=tr^{ }_X \rhohat$ are the reduced density operators.  Clearly, $\mathcal{E}$ is zero for a separable pure state, because in this case $\rhohat^{ }_X$ and  $\rhohat^{ }_Y$ are also projection operators. This is a key feature of a separable state. Since $\mathcal{E}$ measures the deviation of the reduced density operator from Idempotency, it is a measure of the entanglement~\cite{fnote_entangle}. Moreover, $\mathcal{E}=0$ is a necessary and sufficient condition of separability. 

\begin{figure}[t]
   \centering
   \includegraphics[width=7cm]{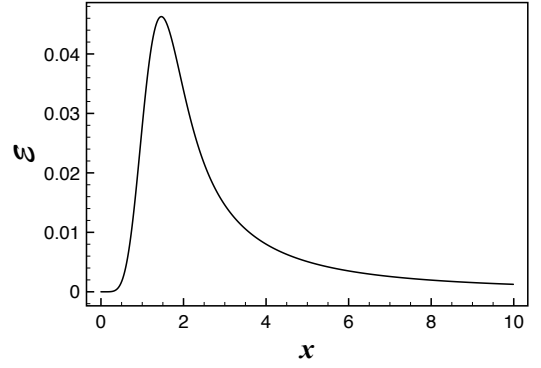} 
   \caption{Entanglement, $\mathcal{E}$, vs. atom-field coupling, $x=\xi/\epsilon$, in the emergent radiation state $|\psi_0\rangle$.}
   \label{fig:entangle}
\end{figure}

Applying this measure to the simplest emergent radiation state $|\psi_0\rangle=e^{-x(\ahat^\dag-\ahat)}|0\rangle$, which in terms of the original radiation and atomic states can be written as: $
|\psi_0\rangle = e^{-\frac{x^2}{2}}\sum_{m=0}^\infty\frac{x^{2m}}{\sqrt{(2m)!}}|m\rangle\otimes\left[|e\rangle-\frac{x}{\sqrt{2m+1}}|o\rangle\right]$, 
we get the following $x$-dependent entanglement in $|\psi_0\rangle$.
\begin{equation}
\mathcal{E}(x)=\frac{1}{2}\left(1-e^{-4x^2}\right)-2\mathcal{A}^2(x)e^{-2x^2},
\end{equation}
where $\mathcal{A}(x)=x\sum_{m=0}^\infty\frac{x^{4m}}{(2m)!}\frac{1}{\sqrt{2m+1}}$. The entanglement is zero for $x=0$, as it should be in the non-interacting case. For weak couplings, $\mathcal{E}$ increases with $x$. However, we find that the entanglement begins to weaken for sufficiently strong $x$ (see Fig.~\ref{fig:entangle}). In fact, $\mathcal{E}\rightarrow 0$ for $x\rightarrow \infty$, because $\mathcal{A}(x)\rightarrow \frac{1}{2}e^{x^2}$, in this limit. This `asymptotic' disentanglement of radiation from atom is a novel cooperative effect, in which the atom lives in a coherent state, $\frac{1}{\sqrt{2}}(|e\rangle-|o\rangle)$ (also noted in the population inversion), supported by an effectively decoupled radiation. Although we have discussed the entanglement properties only in $|\psi_0\rangle$, we expect similar behavior in the higher emergent radiation states, $|\psi_n\rangle$.

To summarize, we have predicted the existence of the free emergent radiation in a system of a two-level atom interacting with a single, twice-resonant mode of the quantized radiation. We have identified a suitable atom-field interaction under which the free-field behavior of the emergent radiation survives, and also established its relation with the interaction in the dipole approximation. Further, we have calculated the time-dependence of the population inversion and the one-photon correlation. We have also reflected upon the experimental meaning this phenomena. The present idea is equally applicable to the systems exhibiting spin magnetic resonance, and may be investigated there. We have also discussed the nature of entanglement in the emergent radiation state of the interacting case.  As an extension of this work, we are developing similar representations for other quantum optical systems. Moreover, one would also like to investigate a model: $H_1=H_0 + \delta \hat{\chi} $, which describes an atom-photon system detuned away from the twice-resonance ($\delta $ being a measure of this detuning).
%
\bibliography{emergent_field}

\begin{thebibliography}{14}
\bibitem[{fno({\natexlab{a}})}]{fnote_lomega}
\bibinfo{note}{One also finds equally spaced eigenstates in the non-interacting
  atom-photon system for $\epsilon=l\omega$, where $l$ is a positive integer.
  In this case, however, the lowest $l$ states are non-degenerate, while the
  rest are two-fold degenerate. It is therefore {\em different} from a free
  radiation. Moreover, the level-spacing in this case is $\omega$, not
  $\epsilon$.}
\bibitem[{fno({\natexlab{b}})}]{fnote_g}
\bibinfo{note}{It is encouraging to note that the free-field behavior of the
  emergent radiation definitely survives to first order in the coupling $g$ in
  $H_{dipole}$, because the leading corrections to the energy eigenvalues of
  $H_{dipole}$ appear in $\calO(g^2)$.}
\bibitem[{fno({\natexlab{c}})}]{fnote_entangle}
\bibinfo{note}{Subsequently, we found that such a measure of entanglement
  already occurs in the literature~\cite{Wootters,Meyer_Wallach}. However, we
  also note that our realization of this measure is very direct and
  uncomplicated.}
\end{thebibliography}
\end{document}